\documentclass[11pt,a4paper]{article}

\usepackage[
  a4paper,
  left=1.55cm,
  right=1.35cm,
  top=1.85cm,
  bottom=1.9cm
]{geometry}

\usepackage{booktabs}
\usepackage{graphicx}
\usepackage{float}
\usepackage{caption}
\usepackage{placeins}
\usepackage[hidelinks]{hyperref}
\usepackage[numbers,sort&compress]{natbib}
\usepackage{amsmath}
\usepackage{amssymb}
\usepackage{microtype}
\usepackage[noblocks]{authblk}

\title{ROSE: Deep Learning exchange-correlation functional\\
from reduced density-matrix representations}

\author[1,2]{Kirill Kulaev}
\author[1]{Alexander Ryabov\thanks{Corresponding author: \texttt{ununbium17@gmail.com}}}
\author[2]{Bogdan Protsenko}
\author[2]{Alexander Guda}
\author[2]{Alexander Soldatov}
\author[1,3]{Evgeny Burnaev}
\author[1]{Vladimir Vanovskiy}

\affil[1]{Skolkovo Institute of Science and Technology, Artificial Intelligence Center,
  Moscow, 121205, Russian Federation}
\affil[2]{Southern Federal University, The Smart Materials Research Institute,
  Rostov-on-Don, 344090, Russian Federation}
\affil[3]{AIRI, Learnable Intelligence Group, Moscow, 121170, Russian Federation}

\date{September 2026}

\begin{document}

\maketitle

\begin{abstract}
We introduce ROSE, a deep-learning exchange--correlation energy functional built
from the occupied-state density matrix.
On GMTKN55 and on additional external benchmarks considered in this work,
ROSE improves over modern neural network functionals trained under matched
protocols and is on average more accurate than, or on par with, most hybrid and
range-separated hybrid functionals considered here.
On the experimental benchmarks considered here it has leading accuracy across
hybrids, range-separated hybrids, and other neural functionals.
These results motivate nonlocal density-matrix representations as a route to
systematic improvement of learned XC functionals with low computational cost.
\end{abstract}

\section{Introduction}

Density functional theory (DFT) is the workhorse of modern quantum
chemistry~\cite{Hohenberg_1964,kohn1965self}, offering a practical balance
between accuracy and cost that makes large-scale modelling of molecules and
materials feasible.
A central ingredient is the exchange--correlation (XC) energy, which collects the
quantum many-body effects beyond the classical Hartree term; because the exact
XC functional is unknown, the accuracy of DFT is largely controlled by the
quality of its approximation.
Traditional semilocal functionals and many neural XC models have been built
primarily on the electron density and related scalar grid quantities
\cite{Skala1,Kirkpatrick2021-tg,Ryabov2020,Ryabov2022,Dick2020-aq,Nagai2020,Zhou2019}.
By the Hohenberg--Kohn theorem the ground-state density formally determines all
ground-state properties, yet the density is only the diagonal of the
one-particle reduced density matrix (1-RDM), whereas exact orbital exchange is a
two-point contraction of that matrix.
The advantage of working with the density matrix is therefore not a claim of
strictly more information than \(\rho\), but a more direct representation of
nonlocal exchange.

Despite decades of progress, reconstructing nonlocal dependence from the density
diagonal and a finite set of local derivatives still yields approximations of
only moderate accuracy and without universal applicability~\cite{Goerigk2017,Mardirossian_2017}.
Persistent failures include underestimated reaction barriers and poor descriptions
of charge transfer, stretched bonds, and dissociation~\cite{Goerigk2017}.
This difficulty is well illustrated by the long history of approximating the
electronic kinetic energy from the density alone: both local and nonlocal
density-based kinetic functionals remain far harder than the elementary
expression available once the occupied-state density matrix is known.
An analogous situation holds for exchange: exact orbital exchange likewise
admits a simple closed form in terms of the 1-RDM, even though evaluating it is
computationally more demanding than semilocal DFAs.
In practice, the most accurate and transferable applied functionals therefore
supplement semilocal dependence with orbital-dependent Hartree--Fock exchange.
That nonlocality, however, brings a steep computational cost: building exact
exchange scales as \(O(N^4)\) with the orbital basis size (before screening),
which limits hybrid DFT for large systems.

In recent years neural network-based XC functionals have seen significant
development.
These efforts include completely new learned functionals as well as studies aimed
at interpolating or improving existing approximations, with attention to
physical constraints and new architectures~\cite{Ryabov2020,Ryabov2022,Nagai2020,Nagai2022,Dick2020-aq,Zhou2019,Kirkpatrick2021-tg}.
Skala~\cite{Skala1} is a particularly strong recent result: trained on a very
large high-accuracy reaction set, it surpasses leading hybrid and range-separated
hybrid functionals on GMTKN55~\cite{Goerigk2017} while retaining semi-local
\(O(N^3)\) cost~\cite{Skala1}.
Architecturally, Skala still starts from meta-GGA features on the
integration grid and augments a multilayer perceptron (MLP) with convolution
layers that learn nonlocality by exchanging information between grid points
and atom-centred coarse features, including equivariant mixing and symmetric
contraction on those atomic channels~\cite{Skala1}.
In that sense the network must \emph{infer} a nonlocal correction from local
ingredients aggregated at atoms, rather than consuming an already nonlocal
electronic representation.
DM21~\cite{Kirkpatrick2021-tg} instead feeds local Hartree--Fock exchange-energy
densities on the grid, together with meta-GGA ingredients, into an
MLP; that makes the cost closer to the high-level methods
used in training.
Geometry optimization with the studied implementation is further
complicated by the cost and numerical sensitivity of finite-difference nuclear
gradients~\cite{Kulaev_2025}.
DeePKS~\cite{DeepKS} reduces onsite projected density-matrix blocks to fixed
scalar descriptors (eigenvalues of onsite density-matrix blocks) and maps them
with an MLP.
Modern equivariant architectures such as MACE~\cite{Batatia2022} show the value
of keeping tensor features inside the network and learning the target
dependency, rather than freezing hand-designed descriptors.
Comparisons between learned exchange--correlation functionals are complicated by
differences in training data and optimization protocols, which obscure the
contributions of architecture and of the training set.
Moreover, incomplete access to training data limits independent reproduction,
fair evaluation of published functionals, and the method development,
including controlled comparison of designs such as
DeePKS~\cite{DeepKS}, DeepHF, Skala~\cite{Skala1}, CIDER~\cite{Bystrom_2022}, and
\(\alpha\)PBE0~\cite{Khan_2025}.

Here we introduce ROSE: the projected occupied-state density matrix is processed
directly by an equivariant graph network that acts as a functional
\(P\mapsto E_{\mathrm{xc}}^{\theta}[P]\).
The network therefore \emph{sees} nonlocal orbital information at the input,
rather than reconstructing it from meta-GGA grid features as in Skala, collapsing
it to onsite scalars as in DeePKS, or requiring Hartree--Fock exchange densities
as in DM21.
The target complexity of a full self-consistent calculation remains \(O(N^3)\)
in the orbital basis size.
We evaluate ROSE alongside competing neural XC architectures trained on a shared,
openly available dataset, with GMTKN55~\cite{Goerigk2017} as the matched test set.
Under these conditions, ROSE outperforms other retrained baselines,
demonstrating improved predictive performance on the same training data and
stronger transferability to experimental benchmarks of different chemical
phenomena, where it is on average the most accurate method considered.

\section{Methods}

\subsection{Kohn--Sham DFT}

Density functional theory determines ground-state properties of interacting
electrons from the electron density \(\rho(\mathbf{r})\) rather than from the
full many-electron wavefunction~\cite{Hohenberg_1964}.
In the Kohn--Sham (KS) formalism~\cite{kohn1965self}, the interacting system is
mapped onto a fictitious system of non-interacting electrons that reproduce the
same ground-state density while moving in an effective potential.
The KS equations are
\begin{equation}
  \left[
    -\tfrac{1}{2}\nabla^{2}
    +
    v_{\mathrm{ne}}(\mathbf{r})
    +
    v_{\mathrm{H}}(\mathbf{r})
    +
    v_{\mathrm{xc}}(\mathbf{r})
  \right]
  \phi_{i}(\mathbf{r})
  =
  \varepsilon_{i}\,
  \phi_{i}(\mathbf{r}),
\end{equation}
where \(\phi_{i}\) and \(\varepsilon_{i}\) are the KS orbitals and orbital
energies, \(v_{\mathrm{ne}}\) is the external (nuclear) potential,
\(v_{\mathrm{H}}\) the classical Hartree potential, and \(v_{\mathrm{xc}}\) the
exchange--correlation potential
\(v_{\mathrm{xc}}(\mathbf{r})=\delta E_{\mathrm{xc}}/\delta\rho(\mathbf{r})\).
The total energy is
\begin{equation}
  E_{\mathrm{tot}}
  =
  T_{\mathrm{s}}
  +
  E_{\mathrm{ne}}
  +
  E_{\mathrm{H}}
  +
  E_{\mathrm{xc}}
  +
  E_{\mathrm{nn}},
\end{equation}
where \(T_{\mathrm{s}}\) is the non-interacting kinetic energy of the KS orbitals,
\(E_{\mathrm{ne}}\) the nuclear--electron attraction,
\(E_{\mathrm{H}}\) the classical Hartree repulsion,
\(E_{\mathrm{nn}}\) the nuclear repulsion, and \(E_{\mathrm{xc}}\) the
exchange--correlation energy.
Exact \(E_{\mathrm{xc}}\) is unknown; practical DFT replaces it by an
approximation.
The familiar local form
\(E_{x}^{\mathrm{LDA}}=-C_{x}\int\rho^{4/3}(\mathbf{r})\,d\mathbf{r}\)
(and GGA/meta-GGA extensions of the type
\(\int f\!\big(\mathbf{r},\rho,\nabla\rho,\ldots\big)\,d\mathbf{r}\))
is a highly successful approximation, but it is not the only admissible
construction of \(E_{\mathrm{xc}}\): it amounts to a substantial LDA
simplification of the intrinsically nonlocal character of exchange.
In this work we retain an explicit density-matrix dependence
\(E_{\mathrm{xc}}[P]\) as a learned approximation.

\subsection{ROSE: neural network density-matrix functional for XC energy}

\begin{figure}[ht]
    \centering
    \includegraphics[width=0.85\linewidth]{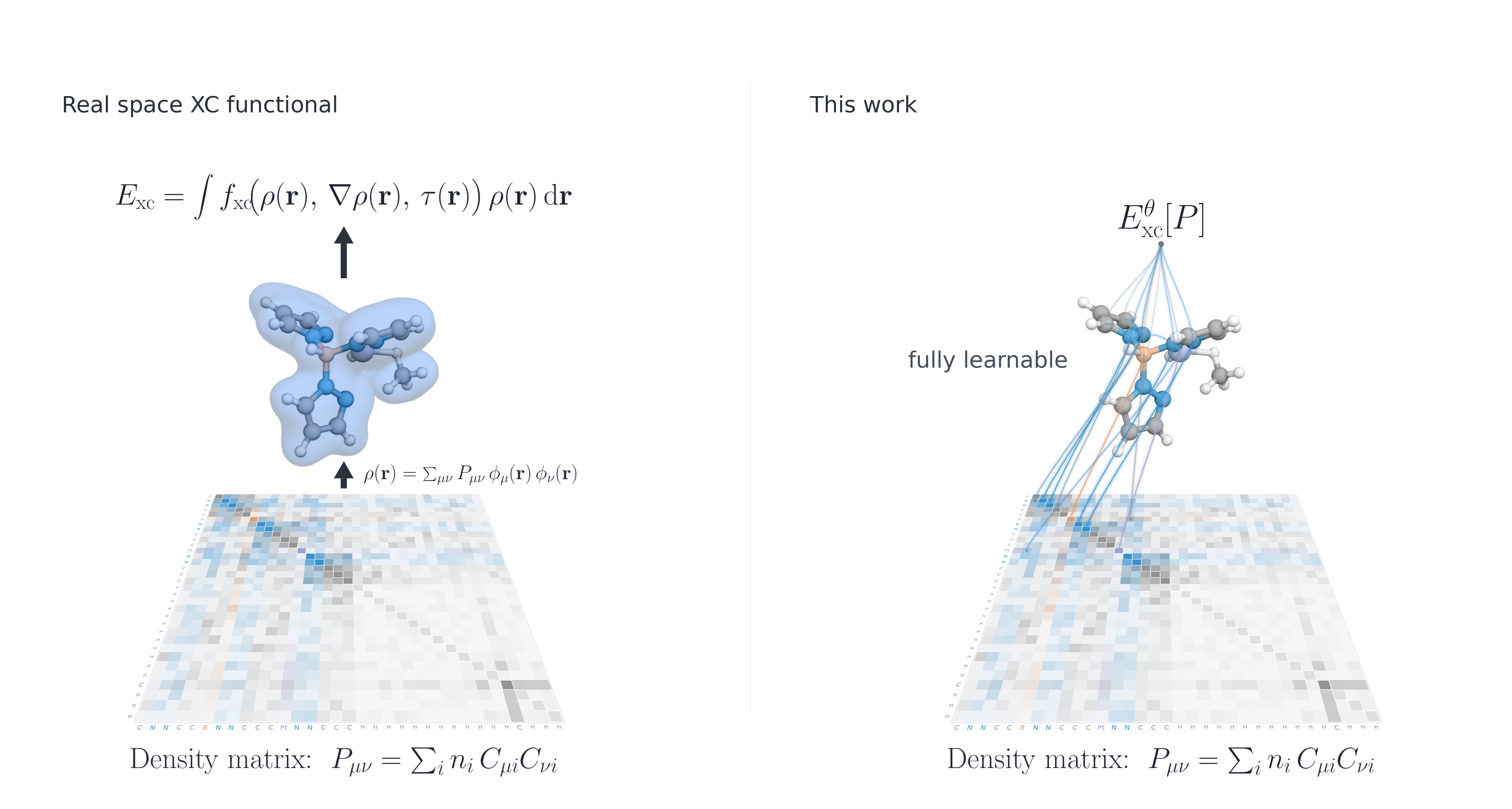}
    \caption{Schematic of ROSE: a neural XC energy functional of the density
    matrix, \(E_{\mathrm{xc}}^{\theta}[P]\) (right), versus a real-space
    density-functional approximation (left).}
    \label{fig:schematic-rho}
\end{figure}

The electronic configuration is represented by the occupied-state density
matrix \(P\), which is processed by a neural network
\(P\mapsto E_{\mathrm{xc}}^{\theta}[P]\).
Technically, the AO density matrix is projected onto a fixed atom-centred
Gaussian projector basis. Onsite blocks of the projected density matrix are
used as node features, while intersite blocks provide edge features.
A message-passing network with MACE symmetric contraction~\cite{Batatia2022}
aggregates onsite and intersite features, and an MLP readout on the
invariant channels yields \(E_{\mathrm{xc}}^{\theta}[P]\).
Full architectural details and hyperparameters are reported in the Supporting
Information (Sec.~\ref{sec:si-architecture}).

The total energy is
\begin{equation}
  E_\mathrm{tot}
  =
  E_{\mathrm{nn}}+
  E_{\mathrm{ne}}+
  T_{\mathrm{s}}
  +
  E_{\mathrm{H}}
  +
  E_{\mathrm{xc}}^{r^2\mathrm{SCAN}}
  +
  E_{\mathrm{xc}}^{\theta}[P]
  +
  E_{\mathrm{D3}}^{r^2\mathrm{SCAN}}
  ,
\end{equation}
where \(E_{\mathrm{xc}}^{r^2\mathrm{SCAN}}\) is the
\(r^2\mathrm{SCAN}\) meta-GGA exchange--correlation
energy~\cite{Furness2020} and \(E_{\mathrm{D3}}^{r^2\mathrm{SCAN}}\) is
Grimme's D3 empirical dispersion correction~\cite{Grimme2010} with
\(r^2\mathrm{SCAN}\)-consistent damping.
Self-consistent ROSE evaluations use the GKS SCF with the nonlocal network
potential derived from \(E_{\mathrm{xc}}^{\theta}[P]\)
(Sec.~\ref{sec:si-scf-potential}).

\subsection{Training and dataset}

\begin{figure}[t]
    \centering
    \begin{minipage}[t]{0.56\textwidth}
        \centering
        \includegraphics[width=\linewidth]{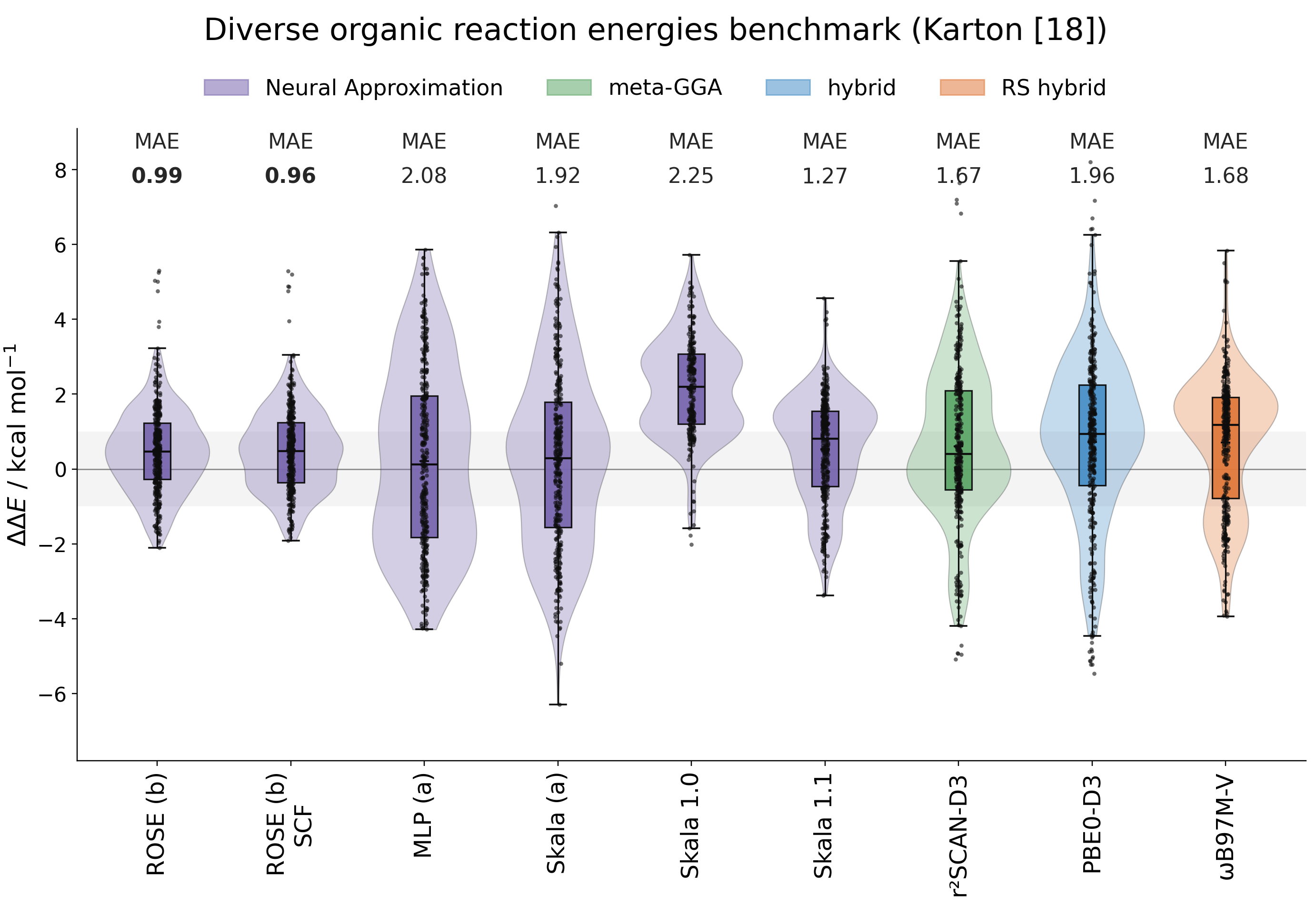}
    \end{minipage}\hspace{0.002\textwidth}%
    \begin{minipage}[t]{0.42\textwidth}
        \centering
        \includegraphics[width=\linewidth]{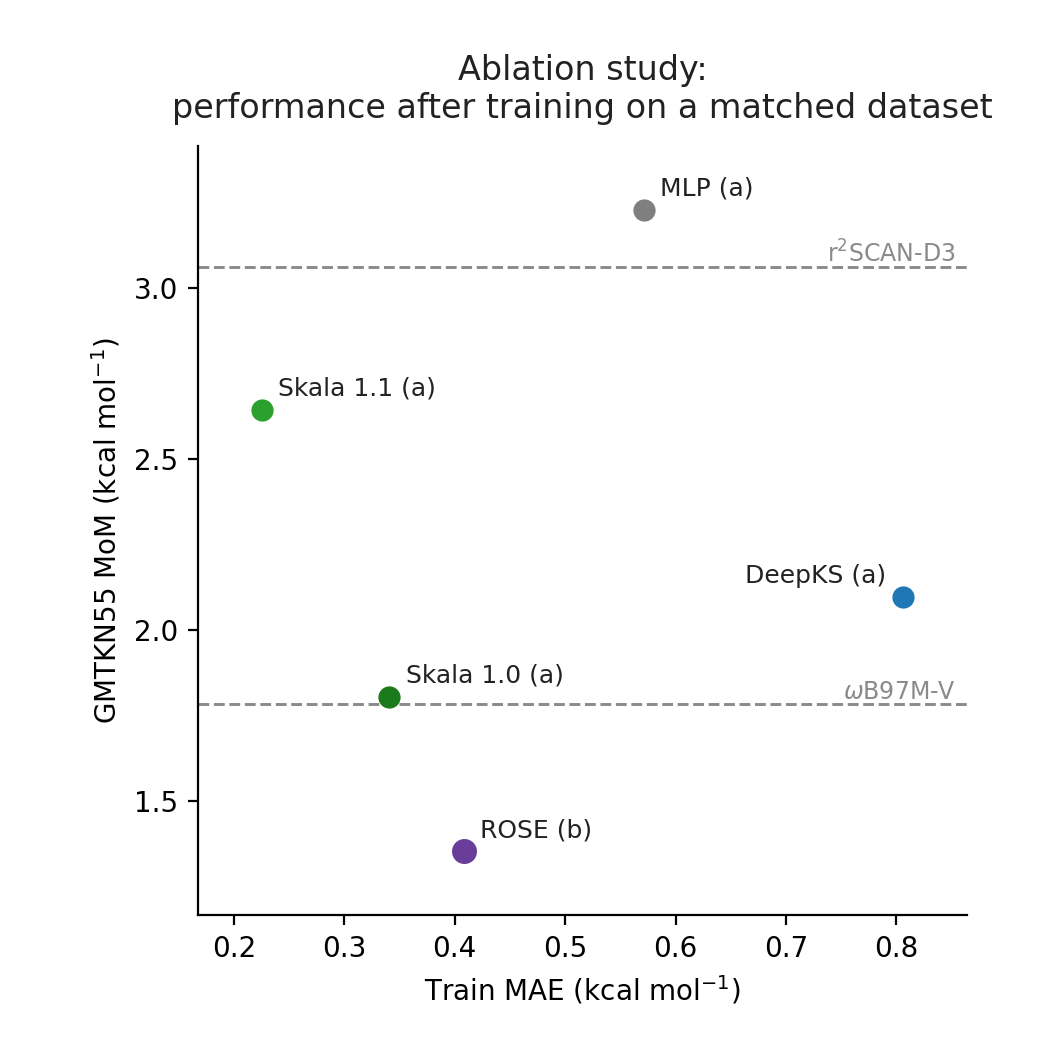}
    \end{minipage}
    \caption{Left: overall error distributions
    (\(\Delta\Delta E\), kcal\,mol\(^{-1}\)) on Karton's high-accuracy test
    dataset of diverse organic reactions~\cite{Karton2026-mm}, including electrocyclic
    reactions, Diels--Alder reactions, electrophilic additions, electrophilic
    aromatic substitutions, and acid--base reactions (calculated with the
    def2-TZVP basis set), with MAE annotated above each method.
    ROSE~(b) evaluated on PBE0 densities (MAE \(0.99\)\,kcal\,mol\(^{-1}\)) and
    in a self-consistent calculation (ROSE~(b)~SCF, MAE \(0.96\)\,kcal\,mol\(^{-1}\))
    is compared with Skala~(a) (\(1.92\)\,kcal\,mol\(^{-1}\)) and MLP~(a)
    (\(2.08\)\,kcal\,mol\(^{-1}\)) trained on the same matched training dataset,
    published Skala~1.0 (\(2.25\)\,kcal\,mol\(^{-1}\)) and Skala~1.1
    (\(1.27\)\,kcal\,mol\(^{-1}\)), and \(\omega\)B97M-V
    (\(1.68\)\,kcal\,mol\(^{-1}\)), alongside classical meta-GGA and hybrid
    functionals.
    Right: ablation on the matched training set---training MAE versus GMTKN55
    mean-of-means (MoM); points are neural XC models trained under
    strategies~(a)/(b) (Sec.~\ref{sec:si-dataset}), with horizontal lines for
    \(r^2\mathrm{SCAN}\) (+\,D3) and \(\omega\)B97M-V (kcal\,mol\(^{-1}\)).}
    \label{fig:matched-pack}
\end{figure}

As discussed in the Introduction, we train all models on the same dataset---GSCDB
(\(4353\) training and \(279\) validation reactions), plus sampled MSR,
RDB7, and DES15k (\(1000\) reactions each) and WATER2510
(\(250\) reactions), altogether \(7603\) reaction
energies per training epoch---so that architectural advantages and the
corresponding transferability can be compared fairly (Fig.~\ref{fig:matched-pack}).
Here each counted element is one reaction energy, not an individual
molecular geometry.
The reaction energy is the stoichiometric combination of total molecular
energies,
\begin{equation}
  \Delta E
  =
  \sum_{M}
  c_{M}\,
  E_{\mathrm{tot}}[M],
\end{equation}
where \(c_{M}\) are the stoichiometric coefficients of species \(M\)
(negative for reactants, positive for products) and \(E_{\mathrm{tot}}[M]\)
is the total energy of molecule \(M\) from the expression above
(with \(E_{\mathrm{xc}}^{\theta}[P]\) evaluated on the fixed training density
of that molecule).
The supervised target is the corresponding high-level reference
\(\Delta E_{\mathrm{ref}}\), and the training loss is a function of
\(\Delta E-\Delta E_{\mathrm{ref}}\)
(Sec.~\ref{sec:si-dataset}).
Full composition, filters, the two training strategies, and optimizer settings
are reported in the Supporting Information (Sec.~\ref{sec:si-dataset});
strategy~(a) follows the original Skala loss function and optimizer protocol of
Ref.~\cite{Skala1}, while
strategy~(b) uses an MSE weighted on the absolute deviations on different training
subsets (a weighting strategy used, for example, for
\(\omega\)B97M-V~\cite{Mardirossian2016}, and which we found to work
slightly better in our setting).

\section{Results}

\subsection{Ablation study results}

We evaluate ROSE against classical DFT protocols with human-crafted XC
functionals, published neural functionals, and neural baselines trained on a
matched dataset
(Skala~(a)/(b), DeePKS~(a)/(b), MLP~(a)) on GMTKN55~\cite{Goerigk2017}.
Skala~(a)/(b) denote Skala retrained on our matched training set under
strategies~(a) and~(b); Skala~1.0/1.1 are published pretrained
models~\cite{Skala1}.
Matched-data columns are labeled accordingly (e.g.\ Skala~(a), ROSE~(b)).

There is material overlap between GMTKN55 and the training set: about
\(14\%\) of GMTKN55 reactions (\(216\) of \(1500\)).
Such overlap is common for widely used benchmarks (for example roughly
\(12\%\) for DM21~\cite{Kirkpatrick2021-tg} and at least \(17\%\) for
\(\omega\)B97M-V~\cite{Mardirossian2016}).
We also report metrics
on subsets whose reactions were not in that overlap, where the same trends hold
(Sec.~\ref{sec:si-dataset}).

Table~\ref{tab:gmtkn55-headline} compares classical functionals and the
published Skala~1.0/1.1 models with ROSE~(b) in a self-consistent
calculation, including WMAD-2 (the metric of Goerigk~\emph{et~al.}~\cite{Goerigk2017})
contributions by reaction type above the
aggregate metrics; ROSE~(b) contributes most strongly on barrier heights and
reaction energies.
In a self-consistent calculation, ROSE~(b) reaches MoM \(1.37\) and
WMAD-2 \(3.15\)\,kcal\,mol\(^{-1}\),
competitive with hybrid and RSH functionals.
Table~\ref{tab:gmtkn55-ablation-mom} reports mean-of-means for matched-training
ablations on full GMTKN55 and on subsets without training overlap; ROSE~(a)/(b)
improve on Skala~(a) (\(1.41\)/\(1.34\) vs \(1.85\)\,kcal\,mol\(^{-1}\) full;
\(1.61\)/\(1.52\) vs \(2.09\)\,kcal\,mol\(^{-1}\) without overlap).
Category-resolved metrics for the matched ablations is given in the Supporting
Information (Sec.~\ref{sec:si-dataset}).

\begin{center}
\footnotesize
\setlength{\tabcolsep}{3.6pt}
\captionof{table}{GMTKN55 metrics (kcal\,mol\(^{-1}\)).
Upper block: WMAD-2 by reaction type; lower block: aggregate MoM / WMAD.
ROSE~(b) is evaluated in a self-consistent calculation; Skala~1.0/1.1 are the published
pretrained models.
All functionals without VV10 corrections use D3(BJ), except M06-2X which uses D3(0)
(as in Ref.~\cite{Skala1}).
All entries use def2-QZVP/ma-def2-QZVP for reactions containing anions.}
\label{tab:gmtkn55-headline}
\resizebox{\textwidth}{!}{%
\begin{tabular}{lrrrrrrrrrr}
\toprule
 & revPBE & r$^2$SCAN & B97M-V & B3LYP & M06-2X & $\omega$B97X-V & $\omega$B97M-V & Skala 1.0 & Skala 1.1 & \textbf{ROSE} (b) \\
\midrule
Basic properties & 1.81 & 1.58 & 1.25 & 1.34 & 0.84 & 1.01 & 0.71 & 0.79 & 0.58 & 0.79 \\
Reaction energies & 1.71 & 1.33 & 1.52 & 1.68 & 0.96 & 1.07 & 0.65 & 0.69 & 0.40 & 0.49 \\
Barrier heights & 1.99 & 1.79 & 0.93 & 1.09 & 0.48 & 0.54 & 0.40 & 0.64 & 0.45 & 0.45 \\
Intermolecular NCI & 1.27 & 1.42 & 0.72 & 1.16 & 1.09 & 0.62 & 0.59 & 0.52 & 0.71 & 0.66 \\
Intramolecular NCI & 1.56 & 1.13 & 1.15 & 1.10 & 1.46 & 0.71 & 0.87 & 1.26 & 0.66 & 0.76 \\
\midrule
MoM & 3.73 & 3.06 & 2.66 & 2.90 & 2.10 & 2.47 & 1.78 & 1.76 & 1.18 & 1.37 \\
WMAD-1 & 4.82 & 4.35 & 3.22 & 3.64 & 2.91 & 2.53 & 2.13 & 2.41 & 1.67 & 1.89 \\
WMAD-2 & 8.34 & 7.25 & 5.56 & 6.38 & 4.83 & 3.96 & 3.23 & 3.89 & 2.80 & 3.15 \\
\bottomrule
\end{tabular}%
}
\end{center}

\begin{center}
\footnotesize
\setlength{\tabcolsep}{5pt}
\captionof{table}{Matched-training ablation: GMTKN55 MoM and WMAD-2
(kcal\,mol\(^{-1}\))
on the full benchmark and on subsets whose reactions were not in the training
overlap.
Evaluated on densities similar to training (PBE0/def2-QZVP;
PBE0/ma-def2-QZVP for anions).}
\label{tab:gmtkn55-ablation-mom}
\begin{tabular}{lrrrrrrrr}
\toprule
 & Skala (a) & Skala (b) & Skala 1.1 (a) & MLP (a) & DeepKS (a) & DeepKS (b) & \textbf{ROSE} (a) & \textbf{ROSE} (b) \\
\midrule
MoM & 1.85 & 2.43 & 2.65 & 3.23 & 2.10 & 2.42 & 1.41 & 1.34 \\
WMAD-2 & 4.00 & 6.51 & 4.46 & 7.51 & 4.90 & 6.07 & 3.22 & 3.07 \\
MoM (no overlap) & 2.09 & 2.69 & 3.05 & 3.67 & 2.39 & 2.71 & 1.61 & 1.52 \\
WMAD-2 (no overlap) & 4.61 & 7.35 & 5.60 & 9.15 & 5.70 & 6.62 & 3.92 & 3.64 \\
\bottomrule
\end{tabular}
\end{center}

\subsection{Transferability to experimental data benchmarks}

\begin{figure}[H]
    \centering
    \includegraphics[width=\linewidth]{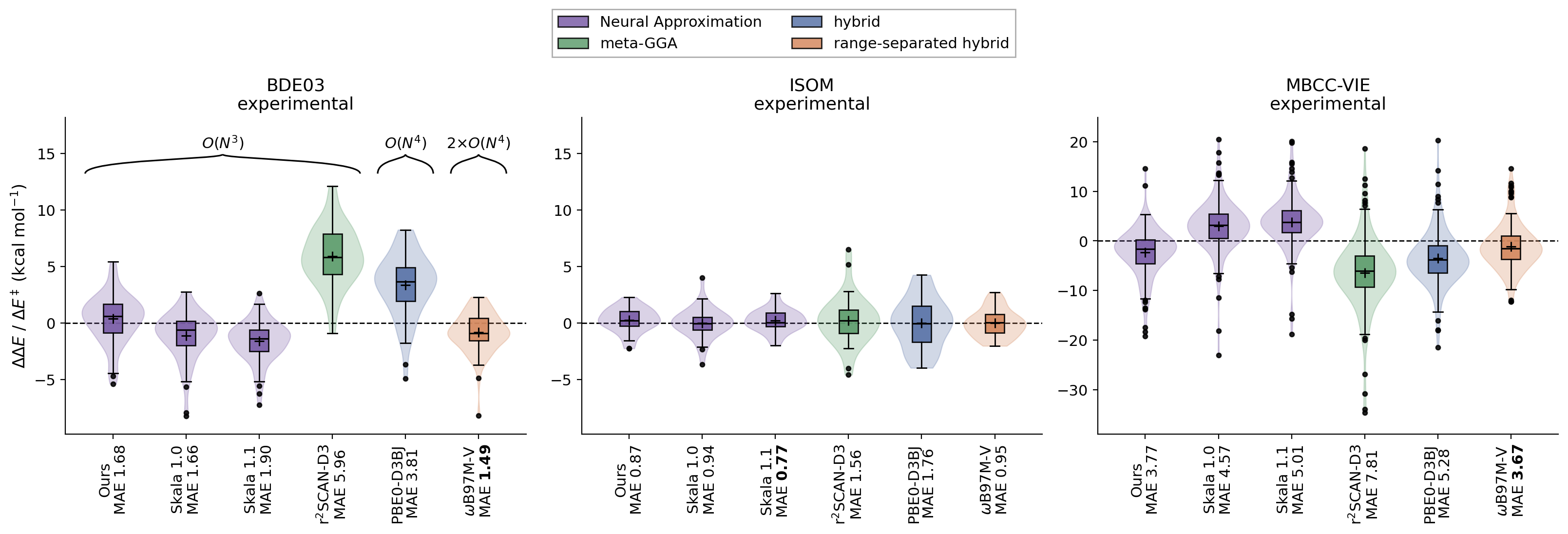}
    \caption{Error distributions on experimental BDE03~\cite{BDE03},
    ISOM~\cite{ISOM}, and MBCC-VIE~\cite{MBCC_VIE} (def2-TZVPPD).}
    \label{fig:benchmarks-ext}
\end{figure}

Figure~\ref{fig:benchmarks-ext} and Table~\ref{tab:ext-exp} summarize errors on
experimental BDE03 (bond dissociation energies)~\cite{BDE03}, ISOM
(isomerization energies)~\cite{ISOM}, MBCC-VIE (vertical ionization
energies)~\cite{MBCC_VIE}, and G3 (atomization energies)~\cite{G3}.
All calculations use the def2-TZVPPD basis; ROSE~(b) is evaluated in a
self-consistent calculation throughout.
As in Table~\ref{tab:gmtkn55-headline}, functionals without VV10 corrections use
D3(BJ), except M06-2X which uses D3(0).

\begin{table}[H]
\centering
\footnotesize
\setlength{\tabcolsep}{2.6pt}
\caption{Experimental benchmarks (MAE, kcal\,mol\(^{-1}\); def2-TZVPPD).
Each method is evaluated in a self-consistent calculation.}
\label{tab:ext-exp}
\resizebox{\textwidth}{!}{%
\begin{tabular}{lrrrrrrrrrr}
\toprule
 & revPBE & r$^2$SCAN & B97M-V & B3LYP & M06-2X & $\omega$B97X-V & $\omega$B97M-V & Skala 1.0 & Skala 1.1 & \textbf{ROSE} (b) \\
\midrule
BDE03~\cite{BDE03} & 5.94 & 6.12 & 2.82 & 3.69 & 2.08 & 1.64 & 1.47 & 1.57 & 1.82 & 1.66 \\
ISOM~\cite{ISOM} & 1.56 & 1.56 & 1.54 & 1.79 & 1.25 & 1.48 & 0.95 & 0.94 & 0.77 & 0.87 \\
MBCC-VIE~\cite{MBCC_VIE} & 10.29 & 7.81 & 6.48 & 5.05 & 3.37 & 3.71 & 3.67 & 4.57 & 5.01 & 3.77 \\
G3~\cite{G3} & 6.96 & 4.11 & 3.25 & 3.45 & 2.15 & 2.32 & 2.83 & 2.28 & 1.61 & 1.63 \\
\midrule
mean MAE & 6.19 & 4.90 & 3.52 & 3.50 & 2.21 & 2.29 & 2.23 & 2.34 & 2.30 & 1.98 \\
mean RMSE & 7.81 & 5.97 & 4.67 & 4.55 & 3.06 & 2.99 & 2.84 & 3.21 & 2.97 & 2.73 \\
\bottomrule
\end{tabular}%
}
\end{table}

Averaged over these experimental sets, ROSE has the lowest mean error,
with mean MAE \(1.98\)\,kcal\,mol\(^{-1}\), ahead of Skala~1.0/1.1 and of
the computationally heavier \(\omega\)B97M-V RSH
(\(2.23\)\,kcal\,mol\(^{-1}\)).
S12L and zeolite adsorption results for larger noncovalent systems are
reported in the Supporting Information (Sec.~\ref{sec:si-s12l}).
\FloatBarrier

\subsection{Computational performance}

\begin{figure}[H]
    \centering
    \includegraphics[width=1.\linewidth]{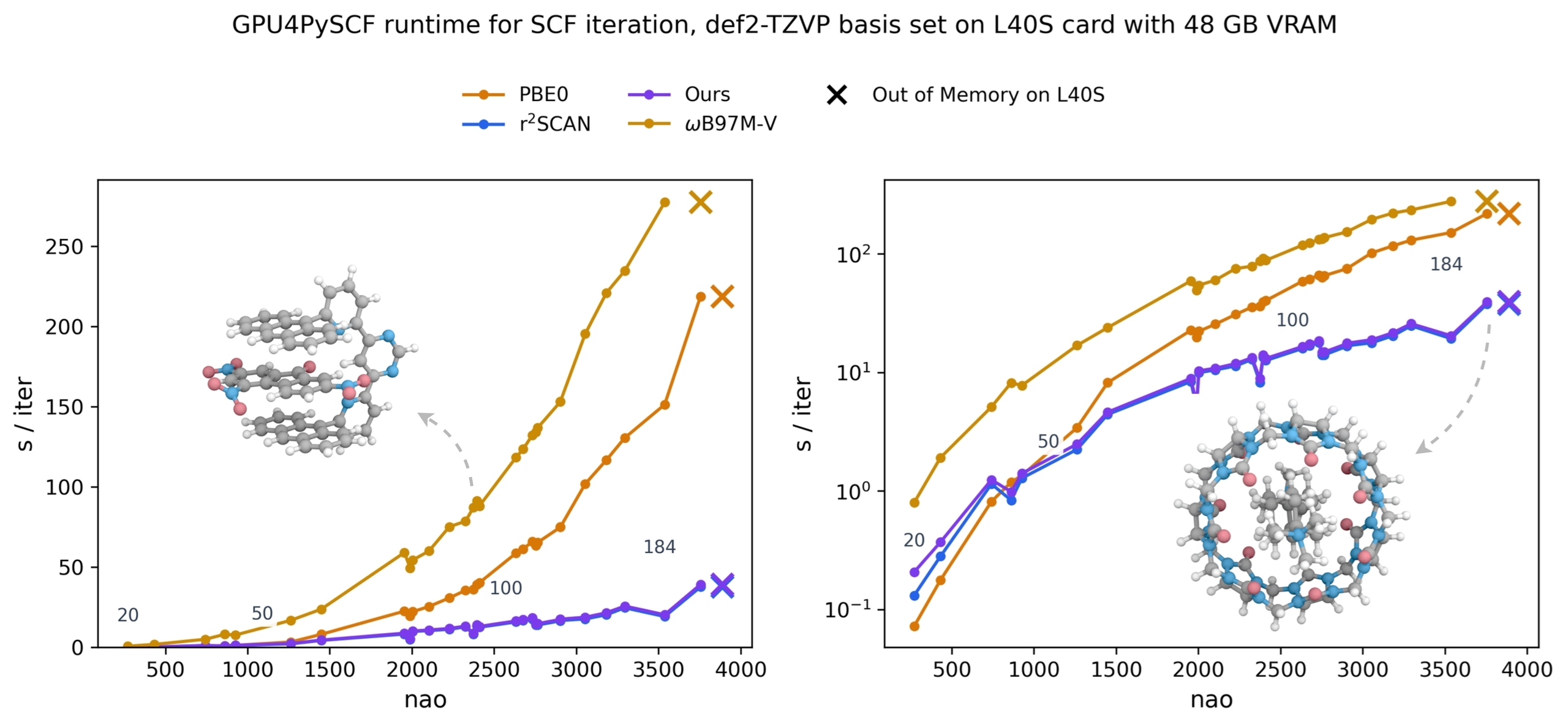}
    \caption{GPU SCF wall time per iteration versus basis size ($N_\mathrm{AO}$)
    for ROSE (Ours), r$^2$SCAN, PBE0, and
    $\omega$B97M-V.
    Calculations use gpu4pyscf~\cite{Li2025-mq} with density fitting
    (def2-TZVP / def2-universal-JKFIT) on an NVIDIA L40S GPU (48\,GB).
    Left: linear scale; right: logarithmic scale.
    Markers show selected system sizes (number of atoms).
    Crosses mark the GPU out-of-memory limit.}
    \label{fig:runtime}
\end{figure}

Figure~\ref{fig:runtime} reports GPU wall time per SCF iteration versus
basis size \(N_\mathrm{AO}\) for ROSE, \(r^2\mathrm{SCAN}\),
PBE0, and \(\omega\)B97M-V.
These timings characterize the cost of one SCF step with the underlying XC
functional under the stated setup.
As seen in the figure, the scaling of ROSE is essentially that of
\(r^2\mathrm{SCAN}\), and on large systems it is an order of magnitude faster
than hybrids and range-separated hybrids.
The molecular systems used for this scaling benchmark are shown in the
Supporting Information (Fig.~\ref{fig:runtime-systems}).

\FloatBarrier

\section{Discussion}

A neural network density-matrix functional can also be useful beyond standard
grid-based Kohn--Sham SCF.
In semi-empirical methods (e.g.\ tight-binding DFT), the
real-space density \(\rho\) is typically never materialized on a quadrature
grid; exchange--correlation and related contributions are instead evaluated as
functionals of the density matrix \(P\) (or of closely related AO-basis
quantities).
The same situation arises in grid-free DFT formulations, where numerical
\(\rho\)-grid quadrature is avoided by construction.
A learned map \(P\mapsto E_{\mathrm{xc}}\) such as ROSE is therefore a natural
candidate for application in those settings.

Incomplete release of training data remains a practical obstacle for independent
reproduction, for evaluating published functionals, and for building alternative
methods under matched conditions.
DM21~\cite{Kirkpatrick2021-tg} does not release its training set, even though
the data construction is described in the paper; for Skala~\cite{Skala1}, about
two-thirds of the reported training reactions are not publicly available,
despite an extensive code release and a companion paper on
MSR-ACC/TAE25~\cite{Skala1}.
Skala's own data ablations further indicate that this unreproducible majority of
specific reaction data is needed to surpass hand-crafted functionals: training on the publicly
accessible thermochemistry-dominated subset alone yields only GGA-level accuracy
on Diet GMTKN55, whereas beating modern hybrids requires the full training
set~\cite{Skala1}.
Our ablation study protocol has its own limits---open training sets are smaller
than Skala's private corpus---so the present study isolates relative data
efficiency rather than claiming an absolute ranking of all published neural XC
models, but GMTKN55 metrics of our trained Skala~1.0 remain close to those
reported in the original work~\cite{Skala1}.
On the electron density benchmark used
in~\cite{Medvedev2017-wv,Mezei2017-jctc,Su2018-pnas,Marjewski2018-mencom,Kirkpatrick2021-tg},
ROSE accuracy is on par with
r$^2$SCAN, \(\omega\)B97M-V, DM21, and Skala~1.0, whereas published
Skala~1.1 is substantially less accurate on this metric and fails to converge
on four highly charged ions (Supporting Information,
Sec.~\ref{sec:si-density}).

\section{Conclusions}

We introduced ROSE, a learned XC functional on the occupied-state density
matrix that maps projected 1-RDM to XC energy.
After matched training, ROSE transfers better than MLP, Skala, DeepKS
models on the GMTKN55 test set.
It also shows highly accurate reaction energies for organic reactions
(MAE below \(1\)\,kcal\,mol\(^{-1}\) on the Karton benchmark of diverse
organic reactions~\cite{Karton2026-mm}).
On the experimental benchmarks considered here
(Fig.~\ref{fig:benchmarks-ext} and Table~\ref{tab:ext-exp}), ROSE has on average
the lowest error among functionals compared, including hybrids,
range-separated hybrids and published Skala~1.0/1.1, while its cost remains that
of semilocal DFT.
Self-consistent density accuracy is comparable to r$^2$SCAN,
\(\omega\)B97M-V, DM21, and Skala~1.0
(Supporting Information, Sec.~\ref{sec:si-density});
per-SCF GPU timings further characterize cost, which is essentially that of
r$^2$SCAN.
Notably, ROSE trained on only about \(7.6\times 10^{3}\) reactions outperforms
published Skala~1.0, which was trained on roughly \(1.5\times 10^{5}\)
reactions~\cite{Skala1}.
Against Skala~1.1 (trained on roughly \(4\times 10^{5}\) reactions and with
a larger model), the comparison is mixed: ROSE is more accurate on average on
the experimental benchmarks considered here (Fig.~\ref{fig:benchmarks-ext}) and
on a dataset of diverse organic reaction energies
(Fig.~\ref{fig:matched-pack}), but less accurate on GMTKN55 aggregates.
These results motivate nonlocal density-matrix representations as a route to
systematic improvement of learned XC functionals.

\FloatBarrier
\bibliographystyle{unsrtnat}
\bibliography{references}

\clearpage
\onecolumn
\begin{center}
{\large\bfseries Supporting Information}\\[0.8em]
{\large ROSE: Deep Learning exchange--correlation functional\\
from reduced density-matrix representations}\\[1.0em]
Kirill Kulaev\textsuperscript{1,2},
Alexander Ryabov\textsuperscript{1},
Bogdan Protsenko\textsuperscript{2},
Alexander Guda\textsuperscript{2},
Alexander Soldatov\textsuperscript{2},
Evgeny Burnaev\textsuperscript{1,3},
Vladimir Vanovskiy\textsuperscript{1}\\[0.6em]
{\small\itshape
\textsuperscript{1}Skolkovo Institute of Science and Technology, Artificial Intelligence Center,
Moscow, 121205, Russian Federation\\
\textsuperscript{2}Southern Federal University, The Smart Materials Research Institute,
Rostov-on-Don, 344090, Russian Federation\\
\textsuperscript{3}AIRI, Learnable Intelligence Group, Moscow, 121170, Russian Federation
}
\end{center}
\vspace{1.2em}

\section{Neural network architecture}
\label{sec:si-architecture}

\paragraph{Projection.}
Let \(\{\chi_\mu\}\) denote the AO basis functions used to expand the KS
orbitals.
Following DeePKS~\cite{DeepKS}, we introduce a fixed set of
atom-centred Gaussian projectors \(\{\alpha_p\}\): for each atom, radial
Gaussian shells indexed by \(n\) are placed at the nuclear position for every
angular momentum \(\ell=0,\ldots,\ell_{\max}\).
The same projector basis is used on every atom.
With the overlap \(S_{\mu p}=\langle\chi_\mu|\alpha_p\rangle\), the
spin density matrix is reduced to this projector space by
\begin{equation}
  P_{\theta}^{\sigma}
  =
  S^{\mathsf T}_{\theta}\, P^{\sigma}\, S_{\theta}.
\end{equation}

For atom \(a\) the corresponding block of \(P_{\theta}\) has size
\(n_{\mathrm{proj}}=108\): \(\ell_{\max}=2\) with twelve radial shells
\(n=1,\ldots,12\) per \(\ell\), so
\(n_{\mathrm{proj}}=\sum_{\ell=0}^{2}12\,(2\ell+1)=12\times(1+3+5)=108\).
Between atoms, intersite blocks \(P^{ab}\) enter the message-passing step
below.

\paragraph{Neural network inputs.}
Atomic features are built from onsite blocks \(P^{aa}\), retaining only
matrix elements with matching radial indices \(n=n'\).
We use the e3nn reduced tensor product (RTP), a fixed Clebsch--Gordan
change of basis that decomposes a matrix block into a direct sum of
\(\mathrm{O}(3)\) irreps.
For a symmetric block of angular momentum \(\ell\) the formula is
\(\mathrm{ij{=}ji}\); for a rectangular block with angular momenta
\(\ell_i\) and \(\ell_j\) it is \(\mathrm{ij}\).
Equivariant features are indexed as in MACE~\cite{Batatia2022} by channel
\(k\), irrep \(L\), and component \(M=-L,\ldots,L\).
Here \(\ell\) labels projector angular momentum in \(P\), while \(L\) labels
irreps of the network features after the RTP.
All equivariant linear maps mix only channels of a common \(\mathrm{O}(3)\)
irrep \((L,p)\) and are written \(W_{kL,\tilde{k}}\).
The RTP outputs of all onsite blocks are packed over radial shells \(n\) and
mixed into a shared hidden feature,
\begin{equation}
  h_{aa,kLM}^{\sigma}
  =
  s_{\mathrm{onsite}}
  \sum_{\tilde{k}}
  W_{kL,\tilde{k}}
  \,
  \big(
  \mathrm{RTP}
  \big(P^{aa,\sigma}\big)
  \big)_{\tilde{k}LM},
\end{equation}
with a fixed onsite scale \(s_{\mathrm{onsite}}\).
The two spin channels are concatenated and mixed,
\begin{equation}
  h_{aa,kLM}(\sigma_1,\sigma_2)
  =
  \sum_{\tilde{k}}
  W^{(\mathrm{spin})}_{kL,\tilde{k}}
  \big[
    h_{aa}^{\sigma_1}
    \,\Vert\,
    h_{aa}^{\sigma_2}
  \big]_{\tilde{k}LM}.
\end{equation}
The message-passing trunk is evaluated for both spin orderings
\((\alpha,\beta)\) and \((\beta,\alpha)\); after the last layer the two
trunk features are averaged,
\(h_{aa}=\tfrac12\big(h_{aa}(\alpha,\beta)+h_{aa}(\beta,\alpha)\big)\),
which is invariant under exchange of the spin labels.

\paragraph{Message-passing layers.}
A forward pass consists of \(T\) layers on the directed radius graph of edges
\(a\leftarrow b\) with \(0<R_{ab}=\|R_a-R_b\|<r_{\max}\), where
\(\mathcal{N}(a)\) is the set of neighbours of atom \(a\) within \(r_{\max}\).
For layer indices \(i=0,\ldots,T-1\), node features are first updated by the
MACE symmetric contraction of correlation order \(\nu=3\)~\cite{Batatia2022},
followed by an equivariant linear map and a residual,
\begin{equation}
  \tilde h_{aa,kLM}^{(i)}
  =
  \sum_{\tilde{k}}
  W^{(i)}_{kL,\tilde{k}}
  \big(
    \mathrm{SC}_{\nu=3}^{(i)}
    \big(h_{aa}^{(i)}\big)
  \big)_{\tilde{k}LM}
  +
  h_{aa,kLM}^{(i)}.
\end{equation}
Blocks of \(P^{ab,\sigma_1}\) sharing the same \(\ell\) (all radial pairs
\(n,n'\)) are mapped by \(\mathrm{RTP}_{\mathrm{ij}}\) and an equivariant
linear layer into an edge representation,
\begin{equation}
  x_{ab,kLM}^{(i)}
  =
  \sum_{\tilde{k}}
  W^{(i,\mathrm{edge})}_{kL,\tilde{k}}
  \,
  \big(
  \mathrm{RTP}_{\mathrm{ij}}
  \big(P^{ab,\sigma_1}\big)
  \big)_{\tilde{k}LM}.
\end{equation}
Invariant edge features of \(0e\) width are obtained by a quadratic
contraction within each irrep \(L\),
\begin{equation}
  h_{ab,c}^{(i)}
  =
  \sum_{L}
  \sum_{k\le k'}
  W^{(i)}_{h,c L kk'}
  \,
  \sum_{M}
  x_{ab,kLM}^{(i)}
  \,
  x_{ab,k'LM}^{(i)},
\end{equation}
where \(c\) indexes the resulting invariant channels and each weight
\(W^{(i)}_{h,c L kk'}\) is a scalar.
Neighbourhood messages update the invariant (\(0e\)) channels using
\(\tilde h\), \(h_{ab}^{(i)}\), and fixed radial basis functions
\begin{equation}
  f_{\mathrm{radial}}^{j}(R_{ab})
  =
  \frac{1+\mathrm{erf}(-\kappa_j R_{ab})}{R_{ab}+\varepsilon},
  \qquad
  \kappa_j\in[\kappa_{\min},\kappa_{\max}].
\end{equation}
The MACE polynomial envelope~\cite{Batatia2022}
(\(p=6\)) multiplies the radial basis;
the message is scaled by
\(f_{\mathrm{env}}(R_{ab})=\exp\!\big(-1/(1-R_{ab}/r_{\max})\big)\)
for \(R_{ab}<r_{\max}\) (and zero otherwise),
\begin{equation}
  h_{aa}^{(i+1)}
  =
  \tilde h_{aa}^{(i)}
  +
  s_{\mathrm{intersite}}^{(i)}
  \sum_{b\in\mathcal{N}(a)}
  f_{\mathrm{env}}(R_{ab})\,
  \phi_{\mathrm{msg}}^{(i)}
  \big(
    \tilde h_{bb}^{(i),0e}
    \,\Vert\,
    h_{ab}^{(i)}
    \,\Vert\,
    f_{\mathrm{cut}}(R_{ab})\,f_{\mathrm{radial}}(R_{ab})
  \big),
\end{equation}
where the sum is added only to the \(0e\) channels,
\(\phi_{\mathrm{msg}}\) is a two-layer MLP with \(\mathrm{SiLU}\) nonlinearity,
and \(s_{\mathrm{intersite}}\) is a residual intersite scale.

\paragraph{Readout.}
The network XC correction is the atom sum of an MLP on the invariant
(\(0e\)) channels of the spin-averaged trunk, with all linear biases set to
zero,
\begin{equation}
  E_{\mathrm{xc}}^{\theta}[P]
  =
  \sum_a
  \mathrm{MLP}\big(h_{aa}^{0e}\big).
\end{equation}
The MLP consists of a linear map \(128\to 96\), two hidden layers
\(96\to 96\) with shifted softplus
\(\sigma_0(u)=\log(1+e^{u})-\log 2\), and a final map \(96\to 1\).
The total molecular energy is
\begin{equation}
  E
  =
  E_{\mathrm{nn}}
  +
  E_{\mathrm{ne}}
  +
  T_{\mathrm{s}}
  +
  E_{\mathrm{H}}
  +
  E_{\mathrm{xc}}^{r^2\mathrm{SCAN}}
  +
  E_{\mathrm{D3}}^{r^2\mathrm{SCAN}}
  +
  E_{\mathrm{xc}}^{\theta}[P].
\end{equation}

\paragraph{Hyperparameters.}
\label{sec:si-hyperparams}
The production ROSE model uses \(T=4\) layers with correlation order \(\nu=3\),
cutoff \(r_{\max}=11.34\,a_0\) (\(6\,\text{\AA}\)), radial-basis exponents
\(\kappa_{\min}=0.0529\,a_0^{-1}\) and \(\kappa_{\max}=0.794\,a_0^{-1}\)
(\(0.1\) and \(1.5\,\text{\AA}^{-1}\)), projector angular momentum
\(\ell_{\max}=2\), node hidden irreps
\(128\times 0e+64\times 1o+32\times 2e\), and edge mid irreps
\(4\times 0e+4\times 2e+4\times 4e\).
The message MLP \(\phi_{\mathrm{msg}}\) has hidden width \(32\) with
\(\mathrm{SiLU}\), and the readout MLP has width \(96\).
The radial cutoff \(f_{\mathrm{cut}}\) uses \(p=6\) as in MACE~\cite{Batatia2022};
\(f_{\mathrm{env}}\) is the soft unit step above.
The onsite scale is fixed at \(s_{\mathrm{onsite}}=0.1\); the intersite scales
\(s_{\mathrm{intersite}}^{(i)}\) are learnable and initialized at \(0.05\) in the
first layer and \(0.1\) thereafter.
Onsite RTP features are summed over radial shells \(n\) using fixed unit
coefficients before the equivariant linear map.
The network has \(264\,904\) trainable parameters.

\paragraph{Self-consistent potential.}
\label{sec:si-scf-potential}
Self-consistent ROSE calculations are generalized Kohn--Sham (GKS).
The network XC potential is obtained by automatic differentiation of
\(E_{\mathrm{xc}}^{\theta}\) with respect to the AO density matrix and
Hermitianization,
\begin{equation}
  V_{\mathrm{xc}}^{\theta,\sigma}
  =
  \frac12
  \Biggl(
    \frac{\partial E_{\mathrm{xc}}^{\theta}}{\partial P^{\sigma}}
    +
    \Biggl(
      \frac{\partial E_{\mathrm{xc}}^{\theta}}{\partial P^{\sigma}}
    \Biggr)^{\!\mathsf T}
  \Biggr).
\end{equation}
Each GKS SCF cycle adds \(V_{\mathrm{xc}}^{\theta}\) to the
\(r^{2}\mathrm{SCAN}\) contribution in the Fock-like matrix before
diagonalization.

\section{Training and dataset}
\label{sec:si-dataset}

\subsection*{Ablation study design}

All matched-data models (Skala, MLP, DeePKS, and ROSE) are trained on
the same open reaction-energy training set and evaluated on
GMTKN55~\cite{Goerigk2017} as the matched test set.
MLP is Skala without nonlocal layers.
Trainable parameter counts are kept comparable across models:
ROSE \(264\,904\),
Skala \(276\,001\),
Skala~1.1 \(385\,217\),
MLP \(265\,474\),
and DeePKS \(236\,198\).
DeePKS~\cite{deepks-kit} uses \(108\) shell-eigenvalue descriptors per atom
(\(\ell_{\max}=2\)), with \(\mathrm{GELU}\) and six hidden layers of width
\(206\).
There is material overlap between GMTKN55 and the training set
(\(216\) of \(1500\) reactions, \(\approx 14\%\)), spanning eight subsets
(overlap/size): Amino20x4 (\(75/80\)), BH76 (\(48/76\)), CDIE20 (\(19/20\)),
PArel (\(19/20\)), HAL59 (\(17/59\)), ICONF (\(16/17\)), TAUT15 (\(14/15\)),
CARBHB12 (\(8/12\)).
Metrics labeled ``no overlap'' drop these subsets
(Table~\ref{tab:si-ablation-types-ex}).
Densities \(\rho\) (and, for ROSE, the projected density matrix \(P\)) are taken
from fixed PBE0 calculations (basis sets as below).
Original Skala~\cite{Skala1} learns an enhancement factor of the LDA
exchange energy density,
\(E_{\mathrm{xc}}^{\theta}
=
\int\rho\,s_\theta[\rho]\,\varepsilon_{\mathrm{x}}^{\mathrm{LDA}}\,d\mathbf r\).
For a fair matched-data comparison we keep that same enhancement construction,
but replace the LDA base by the \(r^2\mathrm{SCAN}\) energy densities and add
D3(BJ,\,\(r^2\mathrm{SCAN}\)) as for ROSE:
\begin{equation}
  \varepsilon_{\mathrm{xc}}^{\theta}(\mathbf r)
  =
  s_\theta(\mathbf r)\,
  \bigl(
    \varepsilon_{\mathrm{x}}^{r^2\mathrm{SCAN}}(\mathbf r)
    +
    \varepsilon_{\mathrm{c}}^{r^2\mathrm{SCAN}}(\mathbf r)
  \bigr),
\end{equation}
\begin{equation}
  E
  =
  E_{\mathrm{nn}}
  +
  E_{\mathrm{ne}}
  +
  T_{\mathrm{s}}
  +
  E_{\mathrm{H}}
  +
  \int
  \rho(\mathbf r)\,
  \varepsilon_{\mathrm{xc}}^{\theta}(\mathbf r)
  \,d\mathbf r
  +
  E_{\mathrm{D3}}^{r^2\mathrm{SCAN}},
\end{equation}
evaluated at the PBE0 density.
When \(s_\theta=1\), this recovers pure \(r^2\mathrm{SCAN}\) (+\,D3).
Matched Skala~(a)/(b) and MLP follow this \(r^2\mathrm{SCAN}\)-enhancement
form.

\paragraph{Composition.}
Table~\ref{tab:si-train} lists the reaction counts used in every matched run.
The training mix totals \(7603\) reactions per epoch
(\(4353+1000+1000+1000+250\)); validation is a fixed GSCDB hold-out of
\(279\) reactions after applying the geometry filter below.

\begin{table}[H]
\centering
\caption{Matched training and validation datasets used for all architecture
comparisons.
Validation is GSCDB-only.}
\label{tab:si-train}
\begin{tabular}{llrr}
\toprule
split & source & \(n_{\mathrm{rxn}}\) & role \\
\midrule
train & GSCDB & 4353 & main-group reaction energies \\
train & MSR & 1000 & main-group thermochemistry \\
train & RDB7 & 1000 & reaction energetics \\
train & DES15k & 1000 & noncovalent / dispersion-oriented \\
train & WATER2510 & 250 & water dimer \\
\midrule
train & \textbf{total} & \textbf{7603} & \\
\midrule
val & GSCDB & 279 & model selection \\
\bottomrule
\end{tabular}
\end{table}

Geometries are restricted to at most \(50\) atoms, and systems with \(d\)-block
metals are excluded from training.
Supervised targets are high-level reaction energies
\(\Delta E_{\mathrm{ref}}\) (kcal\,mol\(^{-1}\)).
For every molecule, \(\rho\) and \(P\) are obtained from PBE0/def2-QZVP
calculations for GSCDB, MSR, and RDB7, and from PBE0/def2-TZVPPD for the other
training subsets.

\begin{figure}[H]
    \centering
    \includegraphics[width=0.95\linewidth]{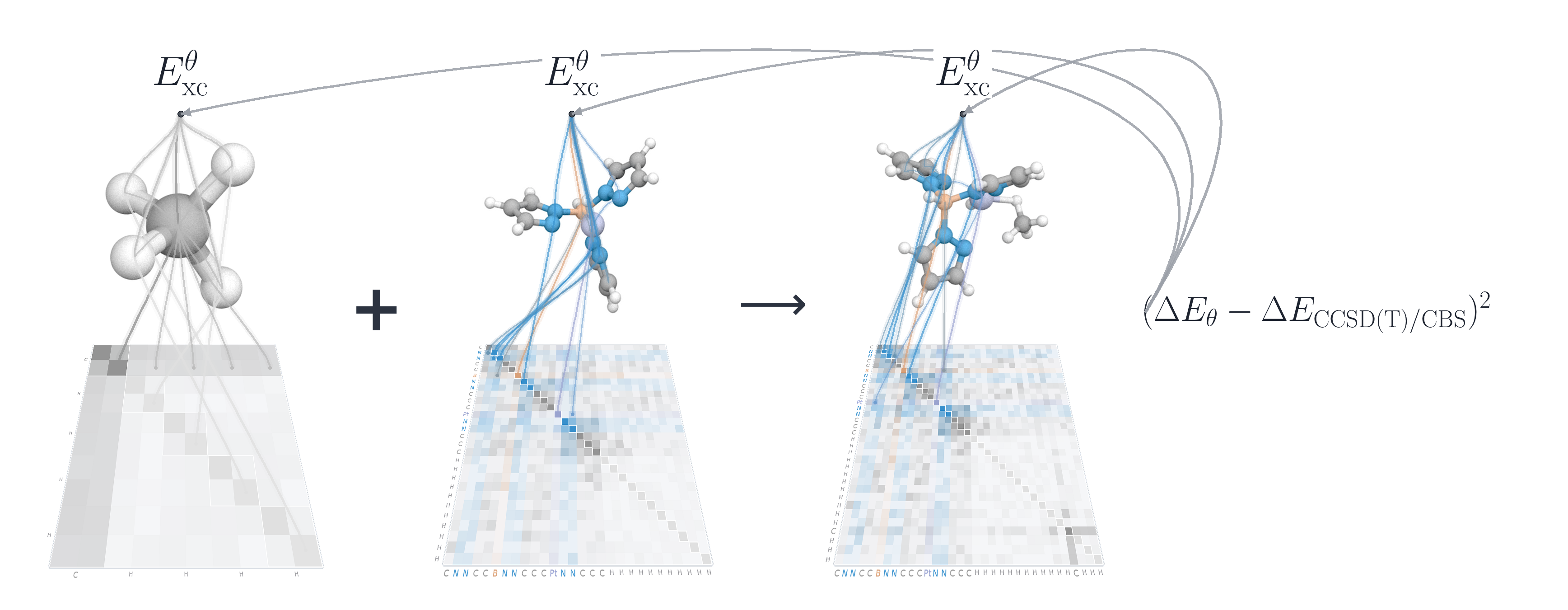}
    \caption{Schematic of the 1-RDM neural XC functional and the reaction-energy
    training protocol.}
    \label{fig:schematic-train}
\end{figure}

We compare two training strategies on this matched training set.

\paragraph{Strategy (a): Skala loss.}
Following Ref.~\cite{Skala1},
\begin{equation}
  \mathcal{L}^{(\mathrm{a})}
  =
  \frac{
    \left|\Delta E-\Delta E_{\mathrm{ref}}\right|^{2}
  }{
    10^{-4}\,E_{\mathrm{h}}
    +
    \left|\Delta E_{\mathrm{ref}}\right|
  },
\end{equation}
with the Muon+Adam optimizer and hyperparameters of that work (Muon peak LR
\(7\times 10^{-4}\), Adam peak LR \(1.5\times 10^{-4}\), Muon momentum 0.95,
Adam \(\beta=(0.9,0.95)\), linear warmup over \(5\times 10^{4}\) steps, cosine
decay over \(3\times 10^{5}\) steps, gradient clipping \(10^{-4}\), EMA decay
0.9999, reaction microbatch size 8).
MLP~(a), Skala~(a), DeePKS~(a), and ROSE~(a) follow strategy~(a).

\paragraph{Strategy (b): MSE weighted on subset baseline errors.}
The reaction loss is
\begin{equation}
  \mathcal{L}^{(\mathrm{b})}
  =
  \left(
    \frac{
      \Delta E_{\mathrm{pred}}-\Delta E_{\mathrm{ref}}
    }{
      \mathrm{MAE}_{\mathrm{base}}
    }
  \right)^{2},
\end{equation}
with one \(\mathrm{MAE}_{\mathrm{base}}\) (kcal\,mol\(^{-1}\)) per training
subset (each GSCDB subset, and MSR, RDB7, DES15k, WATER2510), equal to the
average absolute reaction-energy error of PBE0, M06-2X, \(\omega\)B97X-V, and
\(r^2\mathrm{SCAN}\) on that subset relative to the same high-level references.
Optimization uses AdamW at peak learning rate \(7\times 10^{-4}\), linear warmup
over 5000 steps, cosine decay over \(3\times 10^{5}\) steps, reaction microbatch
size 8, gradient-norm clipping at 0.5, and per-sample loss clipping at 50.
ROSE~(b), DeePKS~(b), and Skala~(b) follow strategy~(b).

\section{Ablation study results}

Table~\ref{tab:si-ablation-types} compares matched-training models by GMTKN55
reaction type
(WMAD-2 contributions).
Table~\ref{tab:si-ablation-types-ex} repeats the comparison on subsets whose
reactions were not in the training overlap.

\begin{table}[H]
\centering
\caption{GMTKN55 WMAD-2 by reaction type for matched-training models
(kcal\,mol$^{-1}$).}
\label{tab:si-ablation-types}
\setlength{\tabcolsep}{4pt}
\small
\begin{tabular}{lrrrrrrrr}
\toprule
type & Skala (a) & Skala (b) & Skala 1.1 (a) & MLP (a) & DeepKS (a) & DeepKS (b) & ROSE (a) & ROSE (b) \\
\midrule
Basic properties & 1.14 & 1.33 & 2.13 & 1.91 & 1.13 & 1.28 & 0.83 & 0.75 \\
Reaction energies & 0.63 & 1.28 & 0.53 & 1.39 & 0.82 & 0.99 & 0.42 & 0.45 \\
Barrier heights & 0.29 & 0.75 & 0.21 & 0.53 & 0.47 & 0.75 & 0.33 & 0.38 \\
Intermolecular NCI & 0.98 & 1.34 & 0.88 & 1.44 & 0.93 & 1.87 & 0.94 & 0.72 \\
Intramolecular NCI & 0.96 & 1.81 & 0.71 & 2.24 & 1.55 & 1.19 & 0.70 & 0.77 \\
\midrule
WMAD-2 & 4.00 & 6.51 & 4.46 & 7.51 & 4.90 & 6.07 & 3.22 & 3.07 \\
MoM & 1.85 & 2.43 & 2.65 & 3.23 & 2.10 & 2.42 & 1.41 & 1.34 \\
\bottomrule
\end{tabular}
\end{table}

\begin{table}[H]
\centering
\caption{Same as Table~\ref{tab:si-ablation-types}, but on GMTKN55 subsets whose
reactions were not in the training overlap (kcal\,mol$^{-1}$).}
\label{tab:si-ablation-types-ex}
\setlength{\tabcolsep}{3.5pt}
\small
\begin{tabular}{lrrrrrrrr}
\toprule
type & Skala (a) & Skala (b) & Skala 1.1 (a) & MLP (a) & DeepKS (a) & DeepKS (b) & ROSE (a) & ROSE (b) \\
\midrule
Basic properties & 1.49 & 1.57 & 2.97 & 2.46 & 1.52 & 1.68 & 1.14 & 1.03 \\
Reaction energies & 0.61 & 1.13 & 0.65 & 1.54 & 1.01 & 1.09 & 0.53 & 0.57 \\
Barrier heights & 0.27 & 0.75 & 0.24 & 0.60 & 0.44 & 0.58 & 0.37 & 0.31 \\
Intermolecular NCI & 1.07 & 1.60 & 0.85 & 1.71 & 0.94 & 2.13 & 1.03 & 0.74 \\
Intramolecular NCI & 1.17 & 2.30 & 0.88 & 2.83 & 1.79 & 1.14 & 0.85 & 0.99 \\
\midrule
WMAD-2 & 4.61 & 7.35 & 5.60 & 9.15 & 5.70 & 6.62 & 3.92 & 3.64 \\
MoM & 2.09 & 2.69 & 3.05 & 3.67 & 2.39 & 2.71 & 1.61 & 1.52 \\
\bottomrule
\end{tabular}
\end{table}

\section{Large systems benchmark}
\label{sec:si-s12l}

Table~\ref{tab:large-systems} reports MAEs on two large-system
noncovalent tests, evaluated self-consistently:
the S12L set of twelve experimental
supramolecular complexes~\cite{Risthaus2013-S12L}
(up to \(177\) atoms in the bound complexes) and zeolite adsorption
energies from Fischer and Brauer~\cite{Fischer_2024}
(DLPNO-CCSD(T)/CBS reference).
For S12L and zeolite/def2-TZVP we also report MAEs after a geometrical
counterpoise (gCP)~\cite{Brandenburg2013} correction, computed with ORCA.
All methods except \(\omega\)B97M-V include their D3(BJ) dispersion
correction.

\begin{table}[H]
\centering
\footnotesize
\setlength{\tabcolsep}{3.5pt}
\caption{MAE on large-system benchmarks (kcal\,mol\(^{-1}\); self-consistent).}
\label{tab:large-systems}
\begin{tabular}{llrrrrrr}
\toprule
 & Basis & ROSE (b) & Skala 1.0 & Skala 1.1 & r$^2$SCAN & PBE0 & $\omega$B97M-V \\
\midrule
S12L & def2-TZVP & 4.40 & 4.90 & 5.00 & 4.85 & 4.74 & 9.56 \\
S12L + gCP & def2-TZVP & 3.35 & 3.30 & 3.19 & 2.97 & 3.13 & 7.69 \\
Zeolite adsorption & def2-TZVP & 2.15 & 10.04 & 7.80 & 4.82 & 5.58 & 4.89 \\
Zeolite adsorption + gCP & def2-TZVP & 0.20 & 7.95 & 5.71 & 2.74 & 3.50 & 2.81 \\
Zeolite adsorption & def2-QZVP & 0.76 & 8.51 & 6.50 & 3.38 & 3.94 & 3.02 \\
\bottomrule
\end{tabular}
\end{table}

\section{Self-consistent electron density assessment}
\label{sec:si-density}

We assess self-consistent densities against all-electron CCSD references for
B\(^{+}\), B\(^{3+}\), C\(^{2+}\), C\(^{4+}\), N\(^{3+}\), N\(^{5+}\),
O\(^{4+}\), O\(^{6+}\), F\(^{5+}\), F\(^{7+}\), Ne\(^{6+}\), Ne\(^{8+}\), and Ne
(basis \texttt{aug-cc-pwCV5Z}),
used for analysis of self-consistent electron densities
in~\cite{Medvedev2017-wv,Mezei2017-jctc,Su2018-pnas,Marjewski2018-mencom,Kirkpatrick2021-tg}.
For each method we evaluate radial distribution errors of the density \(\rho\),
its gradient norm \(|\nabla\rho|\), and its Laplacian \(\nabla^{2}\rho\), with
normalized errors
\[
  \mathrm{NE}_{f}
  =
  \frac{\mathrm{RMSD}[4\pi r^{2}\Delta f]}{\mathrm{median}_{f}},
\]
using fixed median constants for \(f\in\{\rho,|\nabla\rho|,\nabla^{2}\rho\}\),
and
\(\mathrm{MaxNE}
= \max_{\mathrm{systems}}\max(
\mathrm{NE}_{\rho},
\mathrm{NE}_{|\nabla\rho|},
\mathrm{NE}_{\nabla^{2}\rho})\).

\begin{figure}[H]
    \centering
    \includegraphics[width=\linewidth]{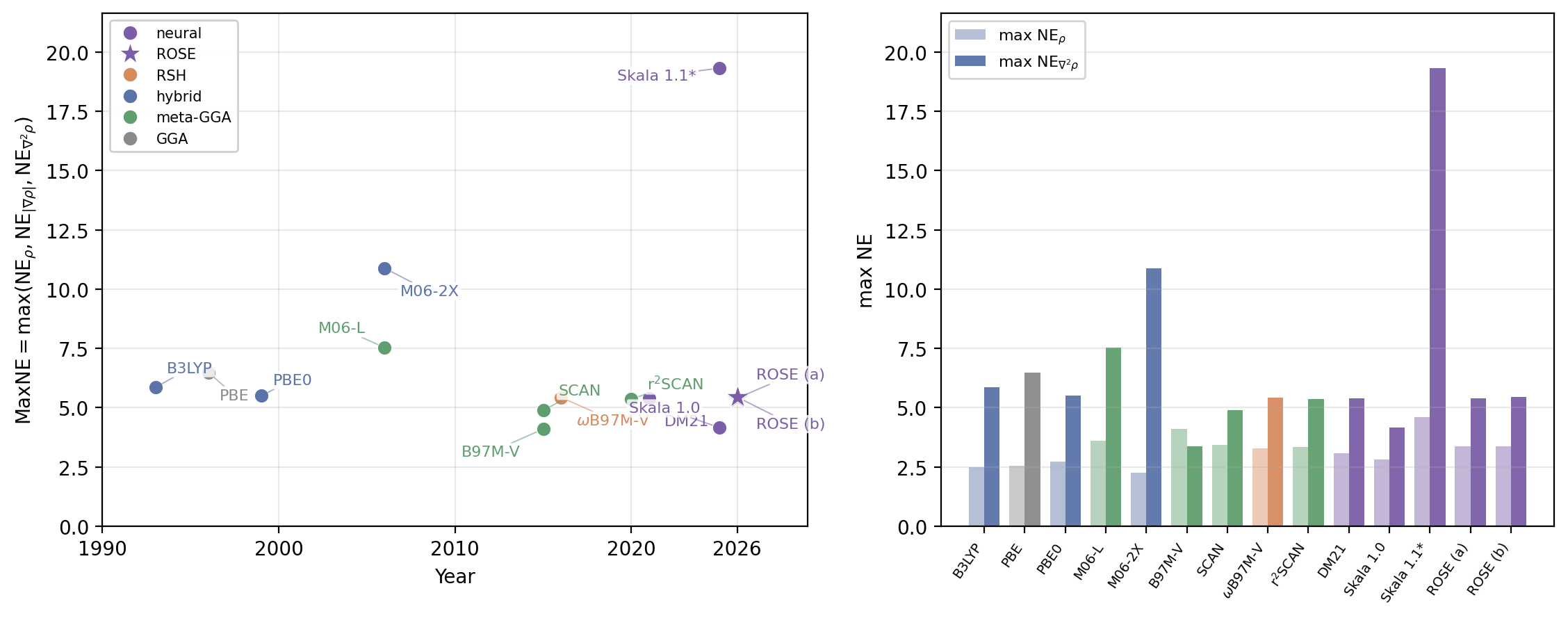}
    \caption{Left: MaxNE versus year for GGA, meta-GGA, hybrid, RSH, and neural
    functionals. Right: breakdown of the maximum
    \(\mathrm{NE}_{\rho}\) and \(\mathrm{NE}_{\nabla^{2}\rho}\).
    $^*$Published Skala~1.1 SCF failed to converge on 4 systems
    (C\(^{4+}\), N\(^{5+}\), O\(^{6+}\), and F\(^{7+}\); energy
    oscillations \(>1\,E_{\mathrm{h}}\), and these systems were not included
    on the plot); MaxNE uses only systems with converged SCF at energy
    tolerance \(10^{-9}\,E_{\mathrm{h}}\).}
    \label{fig:si-density-maxne}
\end{figure}

Figure~\ref{fig:si-density-maxne} places classical and neural functionals by
publication year.

\section{Runtime scaling systems}
\label{sec:si-runtime-systems}

\begin{figure}[H]
    \centering
    \includegraphics[width=\linewidth]{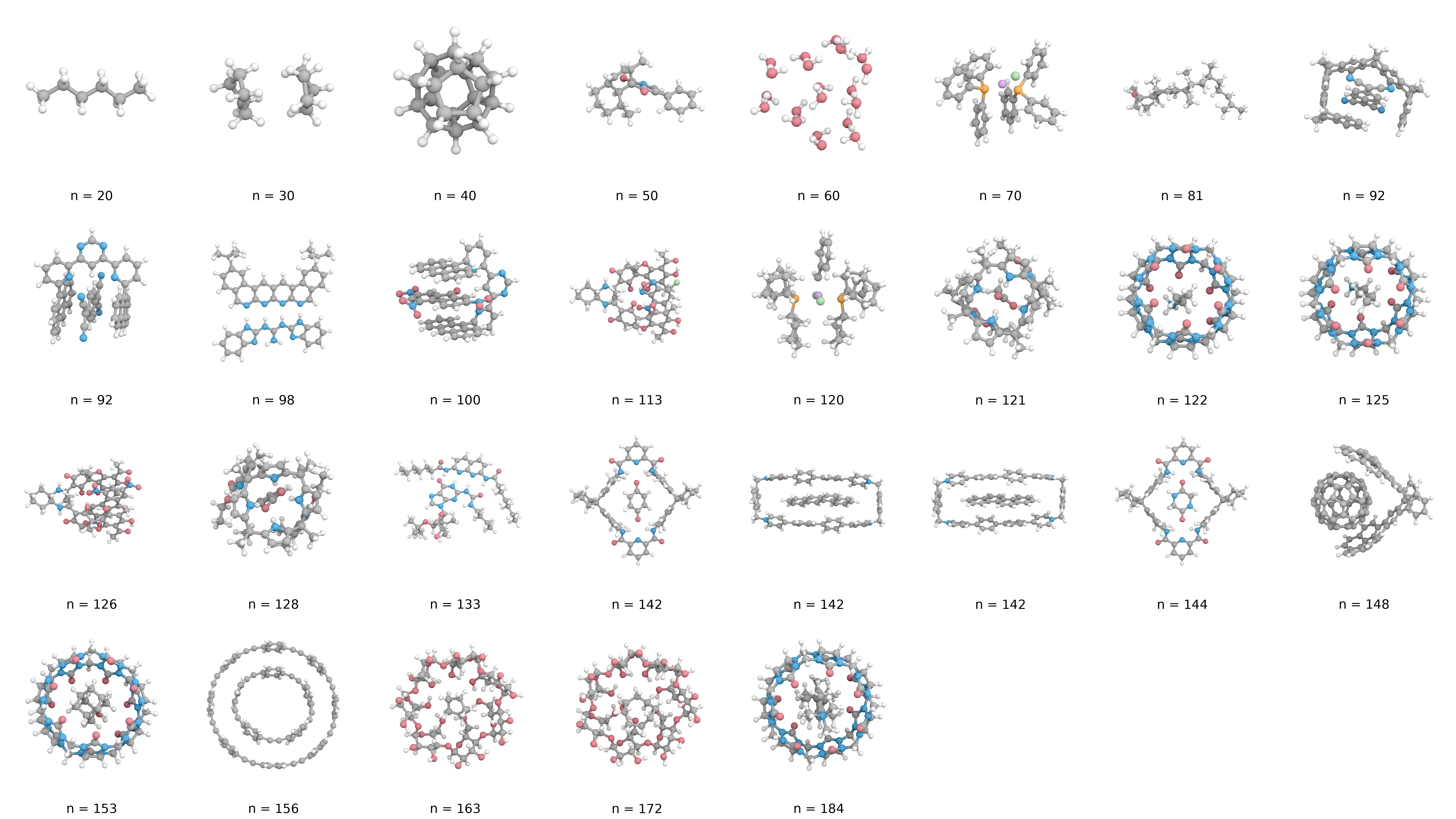}
    \caption{Molecular systems for the GPU SCF wall-time scaling benchmark
    (main text Fig.~\ref{fig:runtime}). Each panel is labeled by atom count \(n\).}
    \label{fig:runtime-systems}
\end{figure}

Figure~\ref{fig:runtime-systems} shows the molecules used for the GPU SCF
timing benchmark in Fig.~\ref{fig:runtime} of the main text.
Labels give the number of atoms \(n\).

\end{document}